\documentstyle[11pt,moriond,epsfig]{article}

\def\Journal#1#2#3#4{{#1} {\bf #2}, #3 (#4)}

\def\PLB{{\em Phys. Lett.}  B}

\def\be{\begin{equation}}
\def\ee{\end{equation}}
\def\bea{\begin{eqnarray}}
\def\eea{\end{eqnarray}}

\begin{document}
\vspace*{4cm}
\title{INFLATION AND THE MICROWAVE BACKGROUND}

\author{Andrew R.~Liddle}

\address{Astronomy Centre, University of Sussex,\\
Brighton BN1 9QJ, Great Britain}

\maketitle\abstracts{I discuss the interplay between inflation and microwave 
background anisotropies, stressing in particular the accuracy with which 
inflation predictions need to be made, and the importance of inflation as an 
underlying paradigm for cosmological parameter estimation.}

\section{Introduction}

Because of calculational simplicity, and because it provides a good fit to 
the observational data, an initial spectrum of adiabatic density 
perturbations is normally assumed responsible for all the observed structures 
in the Universe, such as galaxy clusters and microwave background 
anisotropies. The inflationary cosmology provides a natural explanation for 
such an initial spectrum, and indeed the causal generation of large-scale 
adiabatic perturbations requires a period of inflation.\cite{L95}

In studying the microwave background anisotropies, we believe we have a tool 
through which all of the cosmological parameters, such as the Hubble constant 
$h$ and the density parameter $\Omega_0$, can be probed. This is because the 
evolution of perturbations in the matter and radiation fields depends on all 
the cosmological parameters. On the other hand, the microwave background 
anisotropies are entirely due to linear perturbation theory, and hence 
entirely dependent on the initial perturbations you have in the first 
place. The only reason that one can indeed hope to extract cosmological 
parameters is because one believes that the initial spectrum takes on a 
simple form, perhaps a power-law, which can be parametrized simply and then 
those initial parameters thrown into the melting pot along with the 
cosmological ones and fitted by the data. One of the delights of inflation is 
that the initial spectrum is indeed typically predicted to take on a simple 
form, and furthermore one which can readily be calculated to high precision 
for a given inflationary model.

\section{Models of inflation}

In common with normal practice, I'll focus my discussion on the simplest 
sub-class of inflationary models, where there is a single scalar field $\phi$ 
rolling slowly down some potential $V(\phi)$. The full range of inflationary 
models currently under discussion mostly consists of models which fit into 
this simple class, though I'm keen to stress that there are more complicated 
models on the market, in particular so-called open inflation models which, 
needless to say, give an open Universe. I'll come back to some of these 
complexities later.

The simplest models give rise to a flat spatial geometry, so first let's 
think about how they compare to the observational data. In recent years the 
flood of data supporting a low matter density has become an avalanche, 
leaving even the most hard-bitten theorist feeling guilty to suggest 
otherwise. However, interestingly, the two probes which actually constrain 
the geometry, rather than just the matter density, appear to favour a flat 
geometry. One is the location of the first peak in the microwave background 
spectrum, which is not particularly strongly constrained but certainly looks 
closer to the $\ell = 220$ or so of a flat model than the $\ell = 500$ or so 
of a low-density open model. The second, more secure, evidence comes from the 
recent type Ia supernovae results (Perlmutter et al.\cite{Sup} and Kim in 
these proceedings), which, for a matter 
density of around 0.3 (the favoured value), support the existence of a 
cosmological constant at just the right density to make the Universe flat.

The type of person who likes inflation is rather prone to disliking the 
cosmological constant, on the grounds that it represents exactly the kind of 
fine-tuning which inflation was supposed to liberate us from. On the other 
hand, inflation says nothing about the present matter content of the 
Universe; it just provides a flat spatial geometry and the way in which the 
matter is divided into different classes is, most likely, a property of 
present-day physics. So the presence or otherwise of a cosmological constant 
says nothing about whether inflation really occurred. In fact, one can rescue 
something from this sorry state 
of affairs, by noting that the favoured region corresponds to a Universe 
which is accelerating today. Once people become convinced that the Universe 
is presently accelerating, it will presumably become rather easier to suggest 
that the Universe might also have accelerated at some point in its distant 
past, and as it happens the definition of inflation is precisely any epoch 
during which the Universe experiences accelerated expansion.

\section{Perturbations from inflation}

During an inflationary epoch, comoving length scales are continually being 
stretched to scales larger than the Hubble length. As this happens, quantum 
fluctuations in the fields, which have to exist in accordance with the 
uncertainty principle, become `frozen in' --- unable to evolve on the Hubble 
timescale as their wavelength is so long --- and begin to act like classical 
perturbations. The amplitude of these perturbations is readily calculable, 
and has been studied in many papers. See Liddle and Lyth~\cite{LL93} for a 
review.

Because the Heisenberg uncertainty principle democratically affects 
everything, there are perturbations not only in the scalar field, which 
become adiabatic density perturbations, but also in the gravitational field 
which become gravitational waves. The precise form of both types of 
perturbation, their amplitude 
and scale-dependence, will depend on the potential energy $V(\phi)$, it being 
the only input into the problem.

If the inflationary expansion is sufficiently rapid, those scales on which 
observable perturbations are generated all cross outside the Hubble radius 
during a very narrow interval of time. In that event, physical conditions 
were clearly pretty much the same when the smallest interesting scales 
(corresponding to galaxies) and when the largest interesting scales 
(corresponding to COBE) acquired their perturbations. Hence a first guess is 
that the density perturbations on all scales will be the same, the 
scale-invariant or Harrison--Zel'dovich spectrum. Within this approximation 
all inflation models predict exactly the same outcome, so in particular 
observations could not distinguish between different models.

\begin{figure}
\psfig{figure=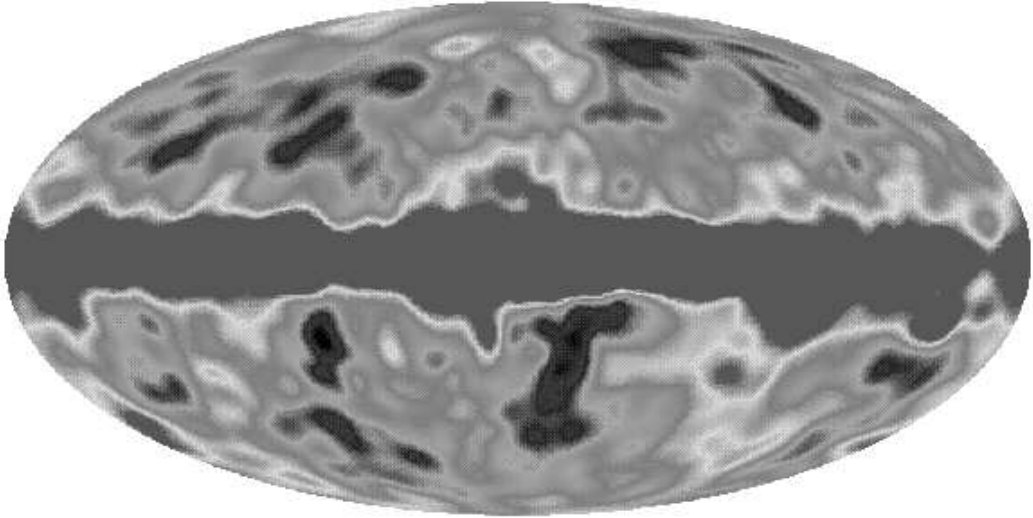,height=7cm}
\caption{One of the maps from the COBE four-year data set.
\label{fig:cobe}}
\end{figure}

Whether this approximation is adequate depends on the data set under 
consideration. Throughout the nineteen eighties, it was advertised as the 
prediction of inflation because existing data, basically the galaxy 
correlation function out to around $10 h^{-1}$ Mpc, covered only a narrow 
range of scales and hence said little about the scale-dependence. 
Interestingly, the COBE data set~\cite{COBE} (seen in Figure 
\ref{fig:cobe}) taken in isolation can also be discussed in this 
approximation. It is consistent with the scale-invariant spectrum and only 
weakly constrains scale-dependence of the spectrum, insufficiently to rule 
out any inflation model on its own.

Fortunately, current data sets, where COBE is combined with short-scale 
information such as the galaxy power spectrum and the galaxy cluster 
abundance, are already of high enough quality that the scale-invariant 
assumption is inadequate, and one must be more precise in assessing the 
inflationary predictions. The next level of sophistication approximates the 
density 
perturbations and gravitational waves as power-laws~\cite{LL92,LL93}; each 
needs an amplitude 
and a spectral index making four numbers in all, and these can readily be 
computed in any given inflation model. In fact there is a degeneracy between 
density perturbations and gravitational waves, the so-called consistency 
equation, which reduces this to three, but in practice it seems optimistic to 
believe that more than one piece of information about the gravitational waves 
can be observationally extracted, and one normally just considers a quantity 
$r$ which measures the fractional contribution of gravitational waves to the 
COBE signal.\cite{LL92,LL93}

Within the power-law approximation different inflation models give different 
predictions for the spectra. This represents a reduction in the 
predictability of inflation as a paradigm, but on 
the other hand means that more information is potentially available to 
observation, yielding extra information on the inflation mechanism and hence, 
perhaps, on very high energy physics.

Although it is often suggested otherwise, the realization that inflation 
doesn't give a perfect scale-invariant spectrum is a huge success for 
the inflationary cosmology. It means that the lowest-order approximation fits 
the data well enough that one has to worry about corrections to it, and that 
the data has reached a quality whereby one can attempt to measure the size of 
these corrections. This is progress of the same sort as occurs in particle 
physics when one realizes that a tree-level calculation is no longer good 
enough and one has to worry about the one-loop corrections.

\section{Scale-dependence of the spectral index}

We're always being told how vastly superior the upcoming microwave background 
observations will be in comparison to the data we presently have, so given 
that present observations already require corrections to the 
Harrison--Zel'dovich spectrum, might we have to worry about correcting the 
corrections? This requires one to consider scale dependence of the spectral 
indices. The optimal strategy appears to be to expand the log of the spectra 
as Taylor series in $\ln k$, i.e.
\begin{equation}
\ln \delta_{{\rm H}}(k) = \ln \delta_{{\rm H}}(k_*) + (n_* -1) \ln
	\frac{k}{k_*} + \frac{1}{2} \, \left. \frac{dn}{d \ln k} 
	\right|_* \ln^2 \frac{k}{k_*} 	\cdots \,.
\end{equation}
Further details can be found in Lidsey et al.\cite{LLKCBA} The expansion 
scale $k_*$ is arbitrary but presumably best chosen in the centre of the 
data. This expansion is to be truncated as soon as an adequate fit to the 
data is obtained. The first term corresponds to the Harrison--Zel'dovich 
spectrum, and the first two taken together to the power-law approximation. To 
these, a 
general inflationary model adds a sequence of derivatives of the spectral 
index, evaluated at the scale $k_*$. In a given model, these are readily 
calculable, in the same way $n$ itself is~\cite{CKLL,KT}. Typically one finds 
that only the first two terms are significant, but there are models where 
higher terms are important too~\cite{CGL,CGKL}.

In general, then, one might want to fit the microwave anisotropy data not 
just for the amplitudes and $n$, but also one or possibly more derivatives of 
$n$. From an inflationary point of view, this looks like a good thing, 
because we are saying that there is an extra piece of information available 
in the microwave anisotropies which we can extract from the data. However, 
there is a downside, which is that the extra piece of information has been 
stolen at the expense of all the other parameters; if we say we need to make 
a fit including one or more extra parameters, then the expected uncertainties 
on {\em all} cosmological parameters will be increased.

We have examined~\cite{CGL} the extent to which the uncertainties are likely 
to increase, using the Fisher information matrix technique~\cite{param}. We 
consider the standard Cold Dark Matter model for illustration; the actual 
numbers aren't very important, what is interesting is the trend as extra 
parameters describing the initial conditions are added. The results are shown 
in Table 1.

\begin{table}[t]
\caption{Estimated uncertainties on parameters expected from the Planck 
satellite, assuming the 
Standard Cold Dark Matter model is correct. Successive columns introduce more 
freedom into the description of the initial parameters. The upper block 
contains the cosmological parameters and the lower one the inflationary 
parameters. Here $\Omega_{{\rm b}}$ is the baryon density, $h$ the Hubble 
parameter, $\Omega_{{\rm \Lambda}}$ a possible cosmological constant and 
$\tau$ the optical depth to the last-scattering surface. $\Omega_{\rm cdm}$ 
is fixed by the assumption of spatial flatness so the second row estimates 
the uncertainty in $h$.\label{tab:param}}
\vspace{0.4cm}
\begin{center}
\begin{tabular}{|l|llll|} 
\hline
Parameter &\hspace{0.5cm}& \multicolumn{3}{c|}{Planck with polarization} \\
\hline 
$\delta \Omega_{{\rm b}} h^2 /\Omega_{{\rm b}} h^2$ & &
 $  0.007$ & $  0.009$ & $ 0.01$ \\
$\delta \Omega_{{\rm cdm}} h^2 /h^2$ & & $0.02$ & $   0.02$ & $ 0.02$ \\
$\delta \Omega_{{\rm \Lambda}} h^2 /h^2$ & & $0.04$ & $   0.05$ & $ 0.05$ \\
$\tau$ & & $ 0.0006$ & $ 0.0006$ & $ 0.0006$\\ 
&&&&\\
$\delta n$ & & $ 0.004$ & $0.04$ & $0.14$ \\
$\delta r$ & & $ 0.04$ & $0.05$ & $0.05$ \\
$dn/d\ln k$ & & $-$ & $  0.006$ & $0.04$ \\
$d^{2}n/d(\ln k)^2$ & & $-$ & $-$ & $0.005$ \\
\hline
\end{tabular}
\end{center}
\end{table}

We assume an experimental configuration of the Planck satellite with 
polarized detectors. The first column shows the results when the power-law 
approximation is assumed, and the successive columns each introduce an 
additional derivative of $n$. First the good news --- the cosmological 
parameters 
take only a very minor hit as extra initial condition freedom is introduced. 
This leads to the encouraging conclusion that the modelling of the initial 
perturbation spectra may not have much of an influence on the satellites' 
abilities to constrain our cosmology. 

The bad news is largely concentrated into the determination of the 
inflationary parameters, and in particular the measurement of $n$ itself, 
whose uncertainty is greatly increased. Note that unless a power-law is {\em 
assumed}, this increase in uncertainty applies even if the values of the 
derivatives are zero to within the observational uncertainties. Nevertheless, 
the loss in accuracy on $n$ may well be overcompensated by the gain in 
information on higher derivatives~\cite{CGKL}; indeed, one might expect that, 
as sneaking in an extra inflationary parameter is a way of transferring a 
small part of the information content in the microwave background away from 
the cosmological parameters and into the inflationary ones.

\section{Inflationary expectations}

As to whether this scale-dependence is likely to show up in practice, we have 
no better guide than the current theoretical prejudice, which says
\begin{itemize}
\item Most slow-roll inflation models do not give a significant 
scale-dependence, even by the standards of Planck.
\item ``Designer'' models of inflation, for example the broken 
scale-invariance models, do give a large effect, but not one which is 
adequately treated within the perturbative framework I've outlined. Such 
models must be confronted with observation on a model-by-model basis.
\item Hybrid inflation models can give an observable effect. Partly this is 
due to the so-called $\eta$-problem; inflation requires that two slow-roll 
parameters $\epsilon$ and $\eta$ be less than one~\cite{LL92,LL93}, but on 
the other hand supergravity models generically predict $\eta = 1 
+$`something'. Since the `something' is unlikely to be extremely good at 
cancelling the 1, such models may well not respect slow-roll very well and 
this enhances the chance of getting detectable scale-dependence. The 
best-motivated models at the moment are those of Stewart.\cite{Ewan}
\end{itemize}

\section{Conclusions}

Inflation as a paradigm is both eminently and imminently testable by upcoming 
microwave background observations. For example, the prediction of a peak 
structure is extremely generic and quite specific to the situation where 
perturbations begin their evolution on scales much larger than the Hubble 
radius, and details such as the peak spacing promises a very strong 
test~\cite{HW}. Something as simple as an observed spectrum without multiple 
peaks appears sufficient to rule out inflation (see e.g.~Barrow and 
Liddle~\cite{BL97}). If inflation passes these tests, then detailed fitting 
to the observations promises startlingly high quality information about the 
inflationary mechanism.

However, the main purpose of my presentation is to provide a reminder of the 
important role inflation has in underpinning the microwave background 
endeavour. I stressed at the start that we can only get highly quality 
constraints from the present radiation power spectrum if we have a simple 
form, preferably motivated by theory, for the initial perturbations. Since 
the observations aim to be accurate at the percent level, the input 
information needs this accuracy too, and inflationary theory is now in a 
position where predictions at this level of accuracy can be made for all 
known models.

This can be contrasted with the situation for topological defect models, 
where it has proven much harder to make accurate theoretical predictions. 
Less accurate theoretical predictions will naturally lead to much more poorly 
determined cosmological parameters. [In fact, Pen (these proceedings) has 
also argued that in a defect model the observed spectrum is less sensitive to 
the cosmological parameters, implying poorer parameter estimation even if the 
theoretical calculations can after all be made more accurate.]

If all goes well with the observations, and inflation proves to be right, we 
can indeed look forward to the tiny error bars one hears about so often. If 
inflation is not correct, the results will still be spectacular but yet after 
the hype the constraints may seem disappointing.

\section*{Acknowledgments}
The author was supported by the Royal Society.

\end{document}